\documentstyle[12pt]{article}
\normalbaselines
\begin{document}

\baselineskip=12pt

\bibliographystyle{unsrt}
\vbox{\vspace{6mm}}

\begin{center}
{\Large Quantum dynamics of a damped deformed oscillator} 
\end{center}

\bigskip 

\begin{center}
Stefano Mancini
\end{center}

\bigskip

\begin{center}
{\it Dipartimento di Fisica and Unit\`a INFM, 
Universit\`a di Milano,\\
Via Celoria 16, I-20133 Milano, Italy}
\end{center}

\bigskip
\bigskip
\bigskip

\begin{abstract}
The interaction of a quantum deformed oscillator with the 
environment is studied deriving a master
equation whose form strongly depends on the type of deformation. 
\end{abstract}

PACS number(s): 03.70.+k, 02.50.-r

\bigskip
\bigskip
\bigskip

Quantum groups \cite{dri} introduced as a mathematical 
description of deformed Lie algebras have
given the possibility to generalize the notion of creation 
and annihiliation operators of the usual
oscillator and to introduce $q$-oscillators \cite{bie,mac}. 
Soon after, several attempts have been
made to give a physical interpretation of the $q$-oscillators. 
In Ref. \cite{mir} they appear in
connection with the relativistic oscillator model, while in 
Ref. \cite{cha} $q$-oscillators have
been used in the generalized Jaynes-Cummings model.

Instead, the nature of $q$-oscillators of the electromagnetic 
field is clarified by the nonlinearity
of the field vibrations \cite{nap}.
This $q$-nonlinearity reflects the exponential growth of 
the frequency of vibrations with the
amplitude.

This observation suggests that there might exist other
types of nonlinearities for which the frequency of oscillation 
varies with the
amplitude via a generic function $f$; this leads to the 
concept of
$f$-deformed oscillators \cite{fde}.
Then, the generation of coherent states for a class of 
$f$-deformed oscillators enter in the real
possibilities of trapped systems \cite{vog}. 

By virtue of possible further physical realizations of 
$f$-oscillators, it could be interesting to
study the environmental effects on such oscillators.

The aim of this paper is to describe the behavior of a 
nonlinear oscillator, like the
$f$-oscillator, plunged in a bath modeled by an assembly 
of harmonic oscillators.
A master equation approach is developed to consider several 
types of reservoir, 
restricted however to the case of small damping.

The free Hamiltonian of an $f$-deformed oscillator may be taken, 
in the case of unitary frequency
and natural units, as \cite{fde}
\begin{equation}\label{Hosc}
{\cal H}_0=\frac{1}{2}\left(A^{\dag}A+AA^{\dag}\right)\,,
\end{equation}
where the operators $A$ and $A^{\dag}$ result as a distortion 
of the usual annihiliation and
creation operators $a$ and $a^{\dag}$,
\begin{eqnarray}
A&=&af(n)=f(n+1)a\,,\quad n=a^{\dag}a\,,\label{Aa}\\
A^{\dag}&=&f(n)a^{\dag}=a^{\dag}f(n+1)\,.\label{Aadag}
\end{eqnarray}
Here, $f$ is intended as an operator-valued function of the 
number operator (moreover it is assumed
Hermitian). In general, it can be made dependent on continous 
parameters, in such a way that, for
given particular values, the usual algebra is reconstructed. 
This is the case of $q$-deformations
\cite{plo}. 
As a consequence of the noncanonical transformations 
(\ref{Aa}) and (\ref{Aadag}), 
the commutation relation is not
preserved\,,
\begin{equation}\label{commu}
\left[A,A^{\dag}\right]=
(n+1)f^2(n+1)-nf^2(n)\,.
\end{equation}
Of course the usual algebra is restored whenever $f\to 1$.
Furthermore, the Eq. (\ref{Hosc}) can be rewritten as
\begin{equation}\label{H0}
{\cal H}_0=\frac{1}{2}\left[nf^2(n)+(n+1)f^2(n+1)\right]\,,
\end{equation}
from which results more clear its nonlinear character. 

Now, the usual model of reservoir consisting in a large 
number of harmonic oscillators is
employed by means of the Hamiltonian ${\cal H}_B
=\sum_j\omega_jb^{\dag}_jb_j$ and the linear
interaction between system and bath \cite{lou}
\begin{equation}\label{Hint}
{\cal H}_{int}=a\Gamma^{\dag}+a^{\dag}\Gamma\,.
\end{equation}
The bath operator $\Gamma$ is given by $\Gamma
=\sum_j\kappa_jb_j$, with $\kappa_j$ the frequency
dependent coupling constants.  
In write down Eq. (\ref{Hint}), we have assumed that 
the frequency of bath modes most strongly
interacting with the system is very large compared to 
the strength of the interaction (measured in
frequency units). This assumption is also known as 
Rotating Wave Approximation \cite{lou}.

We imagine that the interaction (\ref{Hint}) is 
switched on at the time $t=0$ and
the bath at this time have no correlations with 
the oscillator.
The initial joint density operator $w(0)$ can thus 
be taken as the product of some unspecified
operator $\rho(0)$ and a given operator $\rho_B(0)$ 
for the bath. It obeys the Liouville equation
\begin{equation}\label{liuw}
{\dot w}(t)=-i\left[{\cal H}_0+{\cal H}_B
+{\cal H}_{int}\,,\,w(t)\right]\equiv L\,w(t)\,,
\end{equation}
where the superoperator $L$ should be intended as the 
sum $L_0+L_B+L_{int}$. The reduced density
operator of the oscillator will be 
\begin{equation}\label{red}
\rho(t)={\rm Tr}_B\left\{e^{Lt}\,w(0)\right\}
\equiv U(t)\rho(0)\,.
\end{equation}
Assuming the inverse of the operator $U(t)$ to exist, 
we may write $\rho(0)=U(t)^{-1}\rho(t)$, and
thus the formal equation of motion
\begin{equation}\label{liurho}
{\dot\rho}(t)={\dot U}(t)U(t)^{-1}\rho(t)
\equiv{\cal L}(t)\rho(t)\,.
\end{equation}
Restricting ourselves to the weak damping case, 
${\cal L}(t)$ can be constructed perturbatively by
an expansion in powers of the interaction part 
$L_{int}$ of the Liouvillian superoperator $L$.
Hence, neglecting the third and higher order terms, 
the reduced equation of motion is obtained
\begin{eqnarray}\label{rhoeq}
{\dot\rho}(t)=L_0\rho(t)+\int_0^t\,dt'&\Bigg\{&
\langle\Gamma(t')\Gamma^{\dag}\rangle
[a(-t')\rho(t),a^{\dag}]
+\langle\Gamma^{\dag}\Gamma(t')\rangle
[a^{\dag},\rho(t)a(-t')]\nonumber\\
&+&\langle\Gamma(t')\Gamma\rangle
[a^{\dag}\rho(t),a^{\dag}(-t')]
+\langle\Gamma^{\dag}\Gamma^{\dag}(t')\rangle
[a\rho(t),a(-t')]\nonumber\\
&+&{\rm H.c.}\Bigg\}\,,
\end{eqnarray}
where the expectation values are taken on the bath 
and represent its correlation functions.

To go on, it is also necessary the explicit time 
dependence of the operators $a$ and $a^{\dag}$; from
Eq. (\ref{H0}), it is
\begin{equation}\label{aoft}
a(t)=\exp\left[-i\Omega(n)t\right]a
\end{equation}
with
\begin{equation}\label{Om}
\Omega(n)=\frac{1}{2}\left[(n+2)f^2(n+2)-nf^2(n)\right]\,,
\end{equation}
an operator-valued frequency constant in time.

In Eq. (\ref{rhoeq}), by assuming a short correlation 
time of the bath operators
(shorter than any typical system periods),  
the extreme of
integration $t$ can be replaced by $\infty$. Furthermore, 
it is possible to invoke the
fluctuation-dissipation theorem in the following form
\cite{lou}
\begin{eqnarray}\label{fludis}
\int_{-\infty}^{+\infty}dt\,\langle\Gamma(t)\Gamma^{\dag}
\rangle\, e^{i\Omega t}&=&
\chi(\Omega)[1+N(\Omega)]\\
\int_{-\infty}^{+\infty}dt\,\langle\Gamma^{\dag}\Gamma(t)
\rangle\, e^{i\Omega t}&=&
\chi(\Omega)N(\Omega)\\
\int_{-\infty}^{+\infty}dt\,\langle\Gamma\Gamma(t)\rangle\, 
e^{i\Omega t}&=&
\chi(\Omega)M(\Omega)
\end{eqnarray}
where
\begin{equation}\label{chi}
\chi(\Omega)=g(\Omega)\kappa^2(\Omega)\,,
\end{equation}
is the bath response function determined by the density 
function $g$ of the bath modes in the
continuum limit \cite{lou}.
Hence, the master equation (\ref{rhoeq}) becomes
\begin{eqnarray}\label{master}
{\dot\rho}(t)&=&-i[{\cal H}_0,\rho(t)]\nonumber\\
&+&\frac{i}{2\pi}P.V.\int d\Omega'\,\chi(\Omega')\Bigg\{
\left(1+N(\Omega')\right)\left[(\Omega-\Omega')^{-1}a\rho(t), 
a^{\dag}\right]\nonumber\\
&+&N(\Omega')\left[a^{\dag},\rho(t)(\Omega-\Omega')^{-1}a\right]
+M(\Omega')\left[a^{\dag}\rho(t),a^{\dag}(\Omega-\Omega')^{-1}
\right]\nonumber\\
&+&M^{\dag}(\Omega')\left[a\rho(t),(\Omega-\Omega')^{-1}
a\right]-{\rm H.c.}\Bigg\}
\nonumber\\
&+&\frac{1}{2}\Bigg\{\left[\left(1+N(\Omega)\right)
\chi(\Omega)a\rho(t),a^{\dag}\right]
+\left[a^{\dag},\rho(t)N(\Omega)\chi(\Omega)a\right]\nonumber\\
&+&\left[a^{\dag}\rho(t),a^{\dag}M(\Omega)\chi(\Omega)\right]
+\left[a\rho(t),M^{\dag}(\Omega)\chi(\Omega)a\right]
+{\rm H.c.}\Bigg\}\,,
\end{eqnarray}
where $P.V.$ denotes the Cauchy principal value and the 
following relation has been used
\begin{equation}\label{fourier}
\int_0^{\infty}dt\,e^{i\Omega t}{\cal F}(t)=
\frac{1}{2}{\tilde{\cal F}}(\Omega)+\frac{i}{2\pi}P.V.
\int_{-\infty}^{+\infty}d\Omega'\,(\Omega-\Omega')^{-1}
{\tilde{\cal F}}(\Omega')\,,
\end{equation}
with ${\cal F}$ a generic bath correlation function and 
${\tilde{\cal F}}$ its Fourier transform.

Now, the Markov approximation $\chi(\Omega)=2\gamma$ 
\cite{lou}, can be introduced in the second term
with curly brackets in Eq. (\ref{master}), while it is 
inadmissible in the terms involving the
principal value frequency integrals since they become 
divergent. The latter are the analogous 
of the Stark and Lamb shifts in the common theory of 
damping \cite{lou}. 
To simplify the treatment, however, these terms are 
disregarded from now on, due to their small
contribution.

A suitable basis to represent the master equation 
(\ref{master}) in $c$-number form is
obviously the number states basis, then one gets
\begin{eqnarray}\label{masternum}
{\dot\rho}_{m,n}(t)&=&-\frac{i}{2}\Bigg[mf^2(m)+(m+1)f^2(m+1)
-nf^2(n)-(n+1)f^2(n+1)\Bigg]\rho_{m,n}(t)\nonumber\\
&-&\gamma\Bigg\{
m\Big[1+N(\Omega(m-1))\Big]+n\Big[1+N(\Omega(n-1))\Big]
\nonumber\\
&+&(m+1)N(\Omega(m))+(n+1)N(\Omega(n))\Bigg\}\rho_{m,n}(t)
\nonumber\\
&+&\gamma\sqrt{(m+1)(n+1)}\Big[2+N(\Omega(m))+N(\Omega(n))\Big]
\rho_{m+1,n+1}(t)\nonumber\\
&+&\gamma\sqrt{mn}\Big[N(\Omega(m-1))+N(\Omega(n-1))\Big]
\rho_{m-1,n-1}(t)\nonumber\\
&+&\gamma\sqrt{m(n+1)}\Big[M(\Omega(n))+M(\Omega(m-1))\Big]
\rho_{m-1,n+1}(t)\nonumber\\
&+&\gamma\sqrt{(m+1)n}\Big[M^*(\Omega(m))+M^*(\Omega(n-1))
\Big]\rho_{m+1,n-1}(t)\nonumber\\
&-&\gamma\sqrt{m(m-1)}M(\Omega(m-1))\rho_{m-2,n}(t)
\nonumber\\
&-&\gamma\sqrt{n(n-1)}M^*(\Omega(n-1))\rho_{m,n-2}(t)
\nonumber\\
&-&\gamma\sqrt{(m+1)(m+2)}M^*(\Omega(m))\rho_{m+2,n}(t)
\nonumber\\
&-&\gamma\sqrt{(n+1)(n+2)}M(\Omega(n))\rho_{m,n+2}(t)
\,,
\end{eqnarray}
with $\rho_{m,n}(t)\equiv\langle m|\rho(t)|n\rangle$.

It is worth noting how the system deformation strongly 
affects the structure of the master equation
(\ref{masternum}) also in its irreversible terms through 
$\Omega$. Instead, usually the
nonlinearities are included in the master equation only 
through the term $L_0$, so that the
damping terms remain unaffcted by the nonlinearities. 
This seems incorrect, unless 
$M(\Omega)$ and $N(\Omega)$ go to zero.

A further simplification is made by assuming to have only 
a thermal bath, so that
\begin{equation}\label{thermal}
M(\Omega)=0\,;\quad N(\Omega)=\frac{1}{\exp(\beta\Omega)-1}\,,
\end{equation}
where $\beta=1/k_BT$; $k_B$ is the Boltzmann constant and 
$T$ the temperature.
In such a case the stationary solution of the Eq. 
(\ref{masternum}) can be easily found 
\begin{equation}\label{ss}
\rho_{m,n}(\infty)
=\delta_{m,n}\,{\cal Z}^{-1}\,\exp\left[
-\frac{\beta}{2}\left(nf^2(n)+(n+1)f^2(n+1)\right)\right]\,,
\end{equation}
where ${\cal Z}$ is a normalization constant.
The solution (\ref{ss}) may result far from a trivial 
Boltzmann distribution when a deformation is
present; for example, in the case of a (small) 
$q$-type nonlinearity it immediately gives the
deformed Planck distribution derived in Ref. \cite{planap}. 

Finally, it is to remark that the derived master equation 
is not of the Lindblad form \cite{lin},
and this inevitably leads to situations in which unphysical 
density operator will arise
(one illuminating example could be given by the harmonious 
deformation \cite{sud}, for which
$f(n)=1/\sqrt{n}$). 
Nevertheless, this is not a reason to abandon this equation, 
it only means that
one has to be careful in applying it, evaluating 
the range of validity 
for the kind of considered deformation.

Summarizing, a quite general master equation describing the 
damped dynamics of a deformed oscillator has been proposed, 
showing how its structure reflects the nonlinear character 
of the oscillator.
The specific solution strongly depends on the type of 
introduced deformation.
On the other hand, it could be also searched by 
transforming the master equation in a partial
differential equation for a quasiprobability distribution 
\cite{gla}, exploring
possible connection among deformation parameter(s) and 
ordering parameter(s) \cite{gla}.

\section*{Acknowledgments}
I express my gratitude to V. I. Man'ko for introducing me 
to the subject of deformed algebras.

\end{document}